\documentclass[journal=ancac3,manuscript=article]{achemso}
\usepackage[version=3]{mhchem}
\usepackage{epsfig}
\usepackage{amsmath}
\usepackage{color}

\usepackage{fancyhdr}
\newcommand{\T}{\text}

\newcommand{\hl}{\vspace{6pt}\noindent\textbf}

\author{Xuebei Yang}
\altaffiliation{These authors contributed equally to this work.}
\affiliation
{Department of Electrical and Computer Engineering, Rice University, Houston, TX, 77005, USA}
\author{Guanxiong Liu}
\altaffiliation{These authors contributed equally to this work.}
\author{Alexander A. Balandin}
\email{balandin@ee.ucr.edu}
\affiliation
{Nano-Device Laboratory, Department of Electrical Engineering 
and Materials Science and Engineering Program, Bourns College of Engineering, 
University of California-Riverside, Riverside, California, 92521, USA}
\author{Kartik Mohanram}
\email{kmram@rice.edu}
\affiliation
{Department of Electrical and Computer Engineering, Rice University, Houston, TX, 77005, USA}
\alsoaffiliation{Department of Computer Science, Rice University, Houston, TX, 77005, USA}

\title{Triple-Mode Single-Transistor Graphene Amplifier\\and Its Applications}

\begin{document}

\begin{abstract}
In this article, we propose and experimentally demonstrate a triple-mode
single-transistor graphene amplifier utilizing a three-terminal back-gated
single-layer graphene transistor.  The ambipolar nature of electronic transport
in graphene transistors leads to increased amplifier functionality as compared
to amplifiers built with unipolar semiconductor devices.  The ambipolar
graphene transistors can be configured as n-type, p-type, or hybrid-type by
changing the gate bias. As a result, the single-transistor graphene amplifier
can operate in the common-source, common-drain, or frequency multiplication
mode, respectively.  This in-field controllability of the single-transistor
graphene amplifier can be used to realize the modulation necessary for phase
shift keying and frequency shift keying, which are widely used in wireless
applications.  It also offers new
opportunities for designing analog circuits with simpler structure and higher
integration densities for communications applications.

\end{abstract}
\hl{Keywords:} Graphene, Transistor, Ambipolar,
Triple-Mode Amplifier, Phase Shift Keying, Frequency Shift Keying

The single-transistor amplifier, which consists of one transistor and one resistor,
is one of the most basic and most important blocks in analog circuits. 
There are three types of single-transistor amplifiers:
common-source, common-drain, and common-gate, each of which exhibits different
characteristics. The key difference among the three types
of amplifiers is determined by the small-signal voltage gain, defined as $\Delta V_\T{out}/\Delta V_\T{in}$.
The common-source amplifier provides negative gain, whereas the common-drain and common-gate
amplifiers provide positive gain. Since different applications usually prefer different types
of single-transistor amplifiers, it would be very attractive if the same amplifier
can be configured in-field into more than one type. However, in Si-based 
metal-oxide-semiconductor field-effect transistor (MOSFET) technology, the type of an
amplifier is only dependent on its physical configuration, \textit{i.e.},
the node where the input $V_\T{in}$
is applied, the node where $V_\T{out}$ is obtained, and the placement of the resistor. Therefore,
in-field configuration of an amplifier is usually infeasible since the physical configuration
is determined during fabrication.

Recently, graphene, which is a single two-dimensional atomic 
plane of graphite with a honeycomb crystal lattice,
has attracted strong interest as an alternative device technology for future
nanoelectronics~\cite{Novoselov04,Geim07,Li08,Avouris07,Meric08}.
Graphene's high carrier mobility,
excellent mechanical and thermal stability,
superior thermal conductivity~\cite{Balandin08, Ghosh10},
and exceptional resistance to electro-migration make graphene 
an excellent candidate for high-frequency analog applications.
Graphene's high carrier mobility can deliver a large small-signal transconductance $g_m$,
defined as $\Delta I_\T{DS}/\Delta V_\T{GS}$, which is a key parameter determining
the high-frequency performance of a transistor and the gain of an amplifier.
Recent work has demonstrated graphene field-effect transistors with a
cutoff frequency~$f_{T}$ of 100\;GHz~\cite{Lin10} and it has also been predicted that THz
graphene transistors can be achieved at
a channel length of 50\;nm~\cite{Yu09}.
Another important criterion for high-frequency 
analog applications is an acceptably low level of $1/f$ noise. 
It was established that graphene transistors produce relatively 
low levels of $1/f$ noise~\cite{Lin08, Shao09, Liu09}, comparable to those of
conventional semiconductor devices, which makes graphene transistors
suitable for analog applications in terms of their noise spectral density.

The most commonly fabricated graphene transistors use intrinsic micrometer-range graphene layers 
or ribbons as channel material. Owing to the specifics of the band structure of graphene, 
graphene transistors exhibit the ambipolar current conduction behavior. 
In the ambipolar transport regime,  both hole and electron conduction 
are feasible depending on the applied bias~\cite{Geim07}.
By properly adjusting
the gate-source and drain-source voltages, the transistor can be switched from
n-type to p-type, with electron and hole conduction dominating the
current, respectively.
The ambipolar nature of the charge carrier transport may create 
problems for conventional applications based on graphene transistors. 
At the same time, however,
it opens up opportunities for increased functionality in non-traditional circuit architectures. 
For example, graphene transistors have been utilized to demonstrate a frequency
multiplier~\cite{Han09,Wang-apl-10,Wang-nl-10}, a functional logic gate~\cite{Sordan09},
and an inverter~\cite{Harada10}.
However, these designs either focus on the minimum conduction point of the ambipolar curve
where the drain current is at a minimum, which limits the options for design,
or require a four-terminal device
with a top gate and a back gate that are independently controllable,
increasing wiring complexity and operational difficulty.

In this article, we demonstrate a single-transistor amplifier with
three modes of operation utilizing the ambipolarity of a three-terminal graphene transistor.
Depending on whether the graphene transistor is biased at the left branch,
the minimum conduction point, or the right branch of the ambipolar curve,
the amplifier will be configured in the common-drain,
the frequency multiplier, or the common-source mode of operation. 
To the best of our knowledge,
this is the first demonstration of a single-transistor amplifier
that is based on a three-terminal device and that can switch between the common-drain
and common-source modes without altering the physical configuration.
The proposed triple-mode amplifier is demonstrated using a three-terminal back-gated graphene transistor.
We also show theoretically and experimentally that 
our graphene amplifier can greatly simplify communications applications
such as phase shift keying~(PSK) and frequency shift keying~(FSK).
Compared to conventional designs for these applications, the proposed
triple-mode graphene amplifier (i)~has a significantly simpler structure,
(ii)~promises a larger bandwidth and higher frequency of operation,
and (iii)~promises low power consumption.

\begin{figure}
\centering
\includegraphics[width=.9\columnwidth]{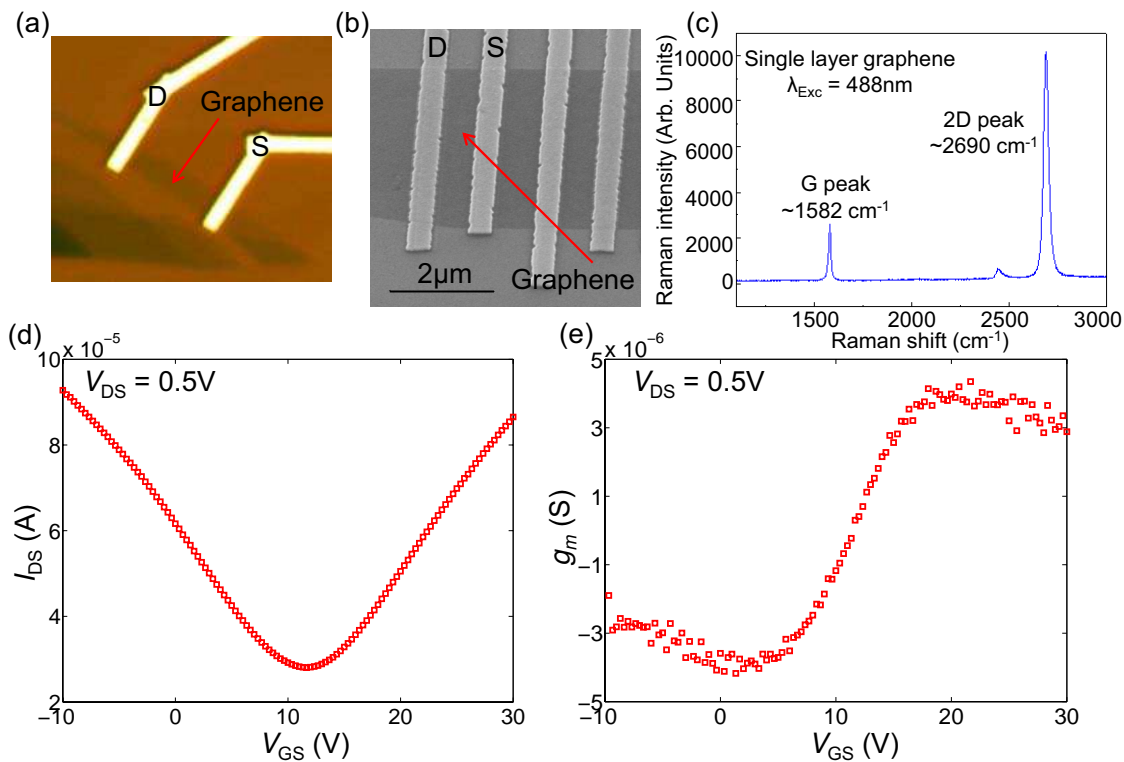}
\caption{
(a)~Optical micrograph image of a representative fabricated back-gated graphene transistor.
(b)~SEM image of source and drain electrodes of a representative back-gated graphene transistor. 
(c)~The Raman spectrum of the single-layer graphene.
(d)~$I_\T{DS}$-$V_\T{GS}$ characteristics of the graphene transistor for $V_\T{DS}=\;$0.5\;V. The current
is minimum at the Dirac point. 
(e)~$g_m$-$V_\T{GS}$ characteristics for $V_\T{DS}=\;$0.5\;V. The transconductance $g_m$
is 0 at the Dirac point.
}
\label{fig:device}
\end{figure}

To demonstrate the triple-mode graphene amplifier, we have fabricated back-gated graphene
transistors from exfoliated graphene flakes.
A representative fabricated device, the scanning electron microscope~(SEM) image,
the Raman spectrum of the single-layer graphene, the $I_\T{DS}$-$V_\T{GS}$ characteristics,
and $g_m$-$V_\T{GS}$ characteristics are shown in~\ref{fig:device}(a), (b), (c), (d) and (e).
Fabrication and measurement details are provided in the methods section at the end of
this article.
Strong ambipolar conduction was observed
in the graphene transistors as evidenced by the ``V''-shaped $I_\T{DS}$-$V_\T{GS}$ curve.
In the ambipolar graphene transistor, the transport is dominated
by electrons and holes for high and low gate voltages, respectively,
and the minimum conduction point $V_\T{min}$ corresponds to the Dirac
point where electrons and holes contribute equally to the transport.
The ambipolar graphene transistor should be regarded
as n-type or p-type at high gate voltage ($V_\T{GS}>V_\T{min}$) or low gate
voltage ($V_\T{GS}<V_\T{min}$), respectively, and as hybrid-type when the gate
voltage is equal to $V_\T{min}$.
The small-signal transconductance $g_m$ is a key factor dominating
the high-frequency performance of a transistor and the gain of the amplifier.
As shown in figure~~\ref{fig:device}(e),
$g_m$ is positive when $V_\T{GS}>V_\T{min}$ and negative
when $V_\T{GS}<V_\T{min}$,
reflecting electron current and hole current, respectively. 

\begin{figure}
\centering
\includegraphics[width=.8\columnwidth]{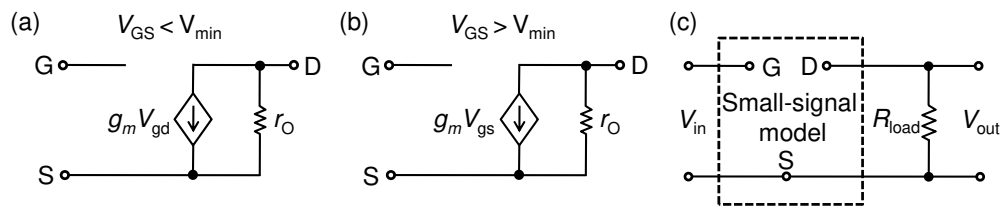}
\caption{
{
%\color{blue}
In (a) and (b), we present the small-signal model for the back-gated graphene
transistor, also referred to as the hybrid-$\pi$ model,
under different $V_\T{GS}$.  Here, $g_m$ is the transconductance
and $r_\T{O}$ is the output resistance.  The small-signal model in~(a)
is used when $V_\T{GS}<V_\T{min}$. Under this condition, the
graphene transistor is p-type and
the small-signal model is similar to that of a p-type MOSFET~\cite{sedra_2004}.
Note that for a p-type MOSFET, the voltage-controlled current source is
controlled by $V_\T{gs}$, yet in the graphene transistor, it is controlled by
$V_\T{gd}$.  This is because in this paper, we always denote the terminal with
higher voltage as the drain for consistency. However, for a p-type MOSFET, the terminal
with higher voltage is usually denoted as the source. Therefore, this
difference arises completely due to the notation used in this paper.
As $V_\T{GS}$ increases, the back-gated graphene
transistor gradually turns from
p-type to n-type and
the small-signal model in~(b) is used when
$V_\T{GS}>V_\T{min}$.
Under this condition, the graphene transistor is
n-type and the small-signal model
is similar to that of an n-type MOSFET~\cite{sedra_2004}.  Note that when
$V_\T{GS}$ is close to $V_\T{min}$, the graphene transistor should be
considered as hybrid-type
instead of either n-type or p-type. Therefore, neither the
n-type nor the p-type small-signal model
is suitable to describe the performance of the graphene transistor.
In~(c), we present the circuit
for small-signal analysis of the triple-mode graphene amplifier from~\ref{fig:OPC}(a).
Note that in small-signal circuit analysis, the power supply is shorted
and the nodes for $V_\T{DD}$ and ground are replaced by a single reference.
}
%$^4$
}
\label{fig:circuit_model}
\end{figure}

The small-signal model for the back-gated graphene transistor,
also referred to as the hybrid-$\pi$ model,
under different $V_\T{GS}$ is shown in~\ref{fig:circuit_model}(a)
and~\ref{fig:circuit_model}(b).  Here, $r_\T{O}$ is the output resistance
and $g_m$ is the small-signal transconductance of the graphene transistor.
Since the graphene transistor is p-type when $V_\T{GS}<V_\T{min}$, the
small-signal model is similar to that of a
p-type MOSFET~\cite{sedra_2004} in~\ref{fig:circuit_model}(a).
Note that for a p-type MOSFET, the voltage-controlled current source is
controlled by $V_\T{gs}$, yet in the graphene transistor, it is controlled by
$V_\T{gd}$.  This is because in this paper, we always denote the terminal with
higher voltage as the drain for consistency. However, for a p-type MOSFET, the terminal
with higher voltage is usually denoted as the source. Therefore, this
difference arises completely due to the notation used in this paper.
Since the transistor is n-type when $V_\T{GS}>V_\T{min}$,
the small-signal model
is similar to that of an n-type MOSFET~\cite{sedra_2004} in~\ref{fig:circuit_model}(b). For $V_\T{GS}$ close to
$V_\T{min}$, the graphene transistor should be considered as hybrid-type
instead of either n-type or p-type. Therefore, neither the
n-type nor the p-type small-signal model
is suitable to describe the performance of the graphene transistor.
Finally,~\ref{fig:circuit_model}(c) illustrates the circuit
for small-signal analysis of the triple-mode graphene amplifier, which will be
introduced in the next section.
Since conventional circuit design has been based on unipolar devices wherein only
one type of carrier dominates the conduction, ambipolar conduction
has usually been considered undesirable. However, our work
is inspired by the ability to leverage the ambipolarity of graphene transistors during circuit
operation.

\section*{Triple-Mode Amplifier}

In this work, we build a triple-mode single-transistor amplifier
using a single back-gated graphene transistor and an off-chip resistor. 
The schematic of the graphene amplifier is shown in~\ref{fig:OPC}(a). The supply voltage $V_\T{DD}$ is set
to 1\;V and the resistor $R_\T{load}$ is 20\;k$\Omega$.
$V_\T{bias}$ is a fixed DC voltage and $V_\T{ac}$ is a small sinusoidal AC signal.
The gate-source voltage of the graphene transistor is hence equal to $V_\T{bias}+V_\T{ac}$.
We show that depending on the relationship between $V_\T{bias}$
and the Dirac point $V_\T{min}$, this amplifier
can have three modes of operation. In each mode, the amplifier exhibits
different performance in 
terms of the small-signal voltage gain $\Delta V_\T{out}/ \Delta V_\T{in}$,
which is given by the expression $\Delta (V_\T{DD}-I_\T{DS}R_\T{load})/ \Delta V_\T{in}$.

\hl{Mode 1 $V_\T{bias}<V_\T{min}$:}
When $V_\T{bias}<V_\T{min}$, the transistor is biased at the left branch of the ambipolar
conduction curve, so
the small-signal transconductance $g_m$ of the transistor
is negative. In the positive phase of $V_\T{ac}$, $I_\T{DS}$ decreases as $V_\T{GS}$
increases. As a result, the voltage drop across the resistor decreases and $V_\T{out}$
increases. It can be similarly inferred that in the negative phase of $V_\T{ac}$,
$V_\T{out}$ will decrease. Therefore, the small-signal voltage gain
in mode 1 is positive, and the input and the output signals have the same phase.
From the transport perspective,
when $V_\T{bias}<V_\T{min}$, the current is mainly due to hole conduction,
so the transistor can be regarded as p-type.
Under this condition, the circuit is configured as a common-drain
amplifier. Analytically, the gain of the amplifier in this mode
is given by the expression $|g_m|R_\T{total}/(|g_m|R_\T{total}+1)$,
where $R_\T{total}$ is the parallel combination of the load resistor $R_\T{load}$
and the inherent output resistance $r_\T{O}$ of the graphene transistor.
This expression can be derived from the small-signal analysis of the complete circuit
illustrated in~\ref{fig:circuit_model}(c),
using the small-signal model for the graphene
transistor shown in~\ref{fig:circuit_model}(a).
The measured results for mode 1 is presented in~\ref{fig:OPC}(c).  The applied
bias voltage $V_\T{bias}$ is 6.5\;V and the frequency of the input AC signal
$V_\T{ac}$ is 10\;kHz.

\hl{Mode 2 $V_\T{bias}>V_\T{min}$:} 
When $V_\T{bias}>V_\T{min}$, the transistor is biased at the right branch of the ambipolar
conduction curve, so
the small-signal transconductance $g_m$ of the transistor is positive.
In the positive phase of $V_\T{ac}$, $I_\T{DS}$ increases as $V_\T{GS}$
increases. As a result, the voltage drop across the resistor increases and $V_\T{out}$
decreases. It can be similarly inferred that in the negative phase of $V_\T{ac}$,
$V_\T{out}$ will increase. Therefore, the small-signal voltage gain
in mode 2 is negative, and the output signal will exhibit a phase shift
of 180$^\circ$ with respect to the input signal. From the transport perspective,
when $V_\T{bias}>V_\T{min}$, the current is mainly due to electron conduction,
so the transistor can be regarded as n-type.
Under this condition, the circuit is configured as a common-source
amplifier. Analytically, the gain of the amplifier in this mode
is given by the expression $-|g_m|R_\T{total}$, where $R_\T{total}$ is
the parallel combination of $R_\T{load}$ and $r_\T{O}$.
As in mode~1, this expression can be derived from the small-signal analysis of
the complete circuit illustrated in~\ref{fig:circuit_model}(c),
using the small-signal model for the graphene
transistor shown in~\protect{\ref{fig:circuit_model}(b)}.
The measured results for mode 2 is presented in~\ref{fig:OPC}(e).
The applied bias voltage $V_\T{bias}$ is 17.5\;V and
the frequency of the input AC signal $V_\T{ac}$ is 10\;kHz.

\hl{Mode 3 $V_\T{bias}=V_\T{min}$:} 
When $V_\T{bias}=V_\T{min}$, the transistor is biased at the minimum conduction point.
In the positive phase of $V_\T{ac}$, the small-signal transconductance is positive.
As a result, the small-signal voltage gain is negative, as analyzed in mode 2.
In contrast, in the negative phase of $V_\T{ac}$, the small-signal
transconductance is negative.  As a result, the small-signal voltage gain of
the amplifier is positive, as analyzed in mode 1.
Thus, when $V_\T{bias}$ is equal to $V_\T{min}$,
the input signal sees a positive gain in its positive phase and a negative gain
in its negative phase, resulting in frequency doubling. The measured
results for mode 3 is presented in~\ref{fig:OPC}(d).
The applied bias voltage $V_\T{bias}$ is 11.1\;V and
the frequency of the input AC signal $V_\T{ac}$ is 4\;kHz.
The spectral purity
of the obtained output was analyzed using the fast Fourier transform.
Frequency doubling effect is clearly observed since
it is observed that 83\% of energy of the output signal is at the frequency of 8\;kHz.
This effect has also been previously reported~\cite{Han09}.

The proposed single-transistor graphene amplifier utilizes the key concept of
biasing in analog circuits,
\textit{i.e.}, only a small range of $I$-$V$ characteristics near the bias point
are necessary to optimize the circuit performance. For this reason, ambipolar
conduction can provide a larger design-space than unipolar conduction because
of the richer diversity of $I$-$V$ characteristics.
Compared to the traditional amplifiers based on
unipolar devices, the proposed single-transistor
amplifier provides greater in-field controllability as 
it can switch between the three modes during operation. To the best of our
knowledge, this is the first work to demonstrate that a single-transistor
amplifier based on a three-terminal device can be in-field configured to
function as both a common-source and a common-drain amplifier.
The small-signal gain observed in the three modes of
operation is $\approx$\;0.01-0.02, which 
is consistent with the small-signal gain that has been reported
for graphene transistors in literature~\cite{Han09,Harada10,Wang-apl-10}.
The low gain can be attributed to the immaturity of the fabrication techniques
common to all graphene devices.
In this article, we demonstrate that the proposed single-transistor triple-mode amplifier
can greatly simplify circuits in common communications
applications such as PSK and FSK.
Both PSK and FSK are important digital modulation
techniques. PSK is widely used in wireless applications such as
Bluetooth, radio-frequency identification (RFID), and ZigBee, while FSK is often used in audio and radio
systems~\cite{xiong_digital_2006}.

\section*{Applications}

\begin{figure}
\centering
\includegraphics[width=.8\columnwidth]{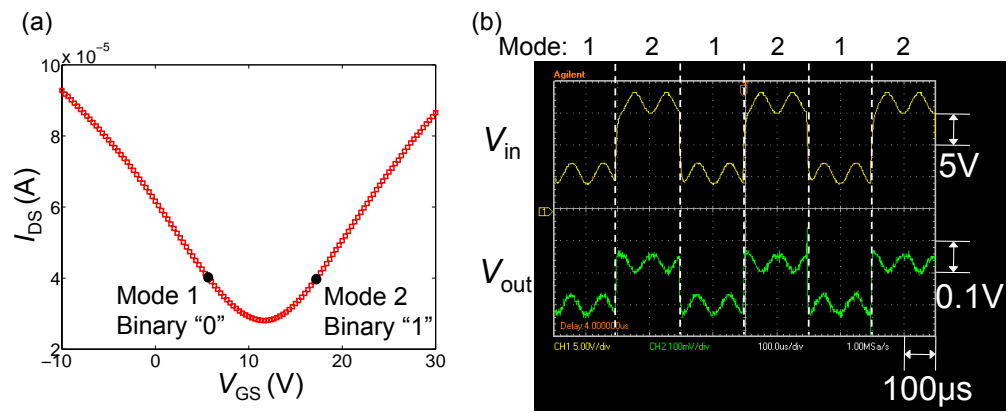}
\caption{
(a)~Two bias voltages, 5.83\;V and 16.8\;V, represent `0' and `1'.
(b)~Experimental results for BPSK modulation. Note that when the bias voltage is 5.83\;V,
the amplifier is configured in mode 1 and the output signal has the same phase as the input signal.
When the bias voltage is 16.8\;V, however, the amplifier is configured in mode 2 and the output
signal has a phase shift of 180$^\circ$ as compared to the input signal.}
\label{fig:PSK}
\end{figure}

We first consider the application of PSK. For brevity, we consider binary~PSK (BPSK)
that is the most basic variant of PSK in this article, but the idea can be extended to
other forms of PSK such as quadrature PSK (QPSK).
In BPSK, the phase of the small AC carrier signal is modulated
and shifted between 0$^\circ$ and 180$^\circ$ to represent the data stream, which takes
the binary value of (0,1).
By using the triple-mode amplifier,
BPSK modulation can be achieved by applying the sinusoid carrier as the small AC signal $V_\T{ac}$
and the data stream, which is the large square wave signal, as the bias $V_\T{bias}$.
If the swing of the square wave signal~$V_\T{bias}$ is chosen such that
the amplifier
can be switched between the positive-gain and negative-gain modes, 
the carrier signal will either experience no phase shift or a phase shift of 180$^\circ$. 
The experimental results for BPSK modulation is presented in~\ref{fig:PSK}.
The biasing voltage $V_\T{bias}$ is switched between
5.83\;V and 16.8\;V, representing digital data `0' and `1', respectively.
It is generated as a square wave signal from
the signal generator. When $V_\T{bias}$ is 5.83\;V,
the graphene transistor is biased at the left branch, so the amplifier
operates in mode~1 with a positive gain. When $V_\T{bias}$ is 16.8\;V,
the graphene transistor is biased at the right branch, so the amplifier
operates in mode~2 with a negative gain. 
The frequency of $V_\T{ac}$ is 10\;kHz. Note that the output signal has different DC voltages
when the amplifier is configured in mode 1 and mode 2, which may not be
preferred during demodulation. However, the DC voltage can be easily filtered
out using a high-pass filter.

\begin{figure}
\centering
\includegraphics[width=.8\columnwidth]{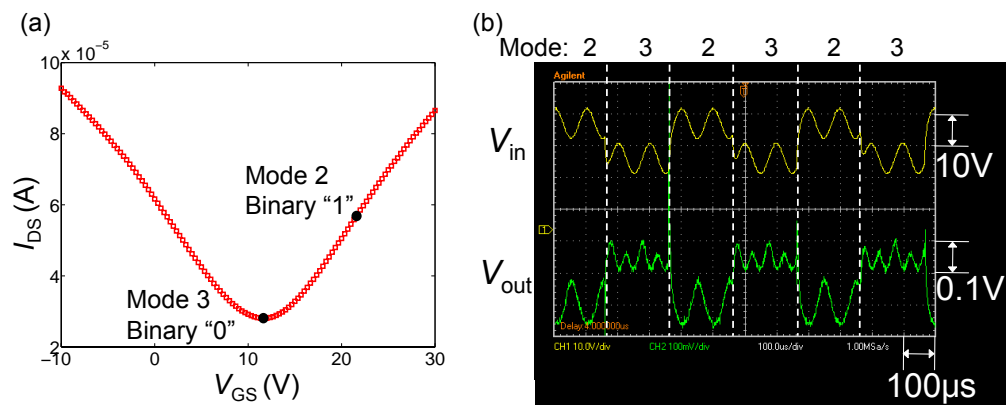}
\caption{
(a)~Two bias voltages, 11.1\;V and 21.9\;V, represent `0' and `1'.
(b)~Experimental results for BFSK modulation. Note that when the bias voltage is 11.1\;V,
the amplifier is configured in mode 3 and the frequency of the output signal 
is doubled in comparison to the input signal.
When the bias voltage is 21.9\;V, however, the amplifier is configured in mode 2 and the output
signal has the same frequency as the input signal.}
\label{fig:FSK}
\end{figure}

We next consider binary FSK (BFSK) that is the most basic variant of FSK for illustration.
In BFSK, the frequency of the small AC carrier signal is modulated
and shifted between $f_\T{c1}$ and $f_\T{c2}$ to represent the data stream, which takes
the binary value of (0,1).
If $f_\T{c2} = 2f_\T{c1}$, such as in the case of Kansas City standard (KCS) for
audio cassette drives where $f_\T{c1}$=1200\;Hz and $f_\T{c2}$=2400\;Hz, BFSK modulation
can be successfully achieved using the proposed triple-mode amplifier. 
Again, as in the case of BPSK,
we can apply the sinusoid carrier as a small AC signal
and the data stream, which is the large square wave signal, as the bias.
If the square wave signal~$V_\T{bias}$ is chosen such that the amplifier is biased 
in mode 3 or in either mode 1/mode 2, the frequency of the
output signal will either be doubled or remain the same, realizing BFSK.
The experimental results for BFSK modulation is presented in~\ref{fig:FSK}.
The biasing voltage $V_\T{bias}$,
generated as a square wave signal from
the signal generator is switched between
11.1\;V and 21.9\;V, representing digital data `0' and `1', respectively.
When $V_\T{bias}$ is 11.1\;V,
the graphene transistor is biased at the Dirac point $V_\T{min}$, so the amplifier
operates in mode 3. When $V_\T{bias}$ is 16.8\;V,
the graphene transistor is biased at the right branch, so the amplifier
operates in mode 2 with a negative gain. The problem of mismatched DC voltage
at the output can be similarly solved by using a high-pass filter.

For comparison, traditional PSK and FSK modulation is usually 
achieved using analog multipliers that
require multiple transistors and/or filtering devices.
However, by leveraging the ambipolar conduction, the proposed amplifier
provides a single-transistor design to achieve PSK
and FSK modulation. It greatly simplifies the circuit design and
the simple structure will potentially also lower power consumption.
Note that the concept described in this article also applies to
other materials exhibiting ambipolar conduction properties, such as 
silicon nanowires~\cite{Koo05}, organic semiconductor
heterostructures~\cite{Dodabalapur95}, and carbon nanotubes (CNTs)~\cite{Heinze03}.
Among these materials, both CNTs and graphene have high mobility that is
preferable for high-frequency analog applications. However, the two-dimensional
planar structure
of graphene enables the current to be easily increased
by increasing the width of the graphene channel, which is advantageous over CNT
transistors.

Given the excellent advantages of the triple-mode amplifier, there are several
directions that merit further investigation to optimize its performance.
Currently, the gain of the amplifier is low and of the order of 0.01-0.02.
This is because (i)~the graphene transistor exhibits low small-signal transconductance $g_m$
and (ii)~the transistor operates in the linear region, with a small inherent
output resistance~$r_\T{O}$.  We believe that this problem can be solved by
improving the device structure and channel quality, increasing the $g_m$,
and pushing the transistor into the saturation region.
Indeed, a frequency multiplier (mode~3 application
of the triple-mode amplifier) with a small-signal gain of~0.15
has been recently reported using the relatively mature carbon nanotube technology~\cite{Wang-nl-10}.
Another challenge is the
mismatch in gain between the different modes for applications such as PSK and
FSK, which may result in extra power loss and higher bit-error rate. We
anticipate that the mismatch will increase as the gain increases.
We believe that the mismatch results from (i) asymmetry in the $I$-$V$
characteristics between the left and the right branch of the ambipolar curve
and (ii)~the inherent performance differences between the common-source and
the common-drain amplifier. Asymmetry between the electron and hole branches can be
reduced by improving the cleanliness of the sample. For example, e-beam resist
residue that is present in the fresh fabricated devices can be removed by
annealing the device in Ar/\ce{H2}~\cite{Ishigami07}.  Inherent
performance differences between the different modes of operation can be reduced by
introducing feedback and using differential outputs.
Other non-idealities in the output signal such as distortions and glitches
exist, but we believe that they can be addressed by improving the quality of
the graphene transistor.

\section*{Conclusions}

In summary, we propose and experimentally 
demonstrate a triple-mode single-transistor graphene amplifier in this article.
The graphene amplifier was built using a three-terminal 
back-gated single-layer graphene transistor and an off-chip
resistor. The ambipolarity of charge 
transport in graphene is an essential element for the triple-mode operation of the amplifier.
Depending on the bias voltage, the amplifier can be configured
in either the common-source, common-drain, or frequency multiplication mode of operation.
To the best of our knowledge, this is the first demonstration of
a single-transistor amplifier that can be tuned between the common-source 
and common-drain configuration using a single three-terminal transistor. 
We also experimentally demonstrated that the in-field controllability 
can be used to realize the modulation necessary for phase shift keying and
frequency shift keying in communications circuits.
As progress is made in
graphene-based thin films for transparent and printable electronics, such
simple circuits deliver both high functionality and in-field configuration
capability necessary for large-scale integration and commercialization.

\section*{Methods}
\hl{Graphene transistor fabrication.} In this work, the back-gated graphene transistor is fabricated using the following methods.
Graphene flakes were placed on a standard silicon substrate with 300\;nm \ce{SiO2} on the top.
The number of atomic layers and quality of graphene flakes were verified
by micro-Raman spectroscopy through the conventional procedure of the 2D/G'
Raman-band deconvolution~\cite{Ferrari06,Calizo07,Calizo-apl-07}. The p-type degenerately doped Si substrate was
used as the back gate to tune the Fermi-level position of graphene.
The source and drain electrodes were fabricated by electron beam lithography (EBL)
followed by the electron beam evaporation of Ti/Au with the thickness of 8/80\;nm.
The channel width of the fabricated devices was~2\;$\mu$m and the length was~9\;$\mu$m.
The DC electrical characteristics of the fabricated graphene transistors were
measured by the probe station (Agilent 4142) under ambient conditions~\cite{Rumyantsev-10}.
The gate biases ranging from -10\;V to 30\;V were applied for the back gate measurements at
a fixed drain bias of 0.5\;V.  The $I_\T{on}$/$I_\T{off}$ ratio was around 3,
while the charge carrier mobility of these devices was in the range
3000--4000\;cm\textsuperscript{2}V$^{-1}$s$^{-1}$ at room temperature.

\begin{figure}
\centering
\includegraphics[width=.95\columnwidth]{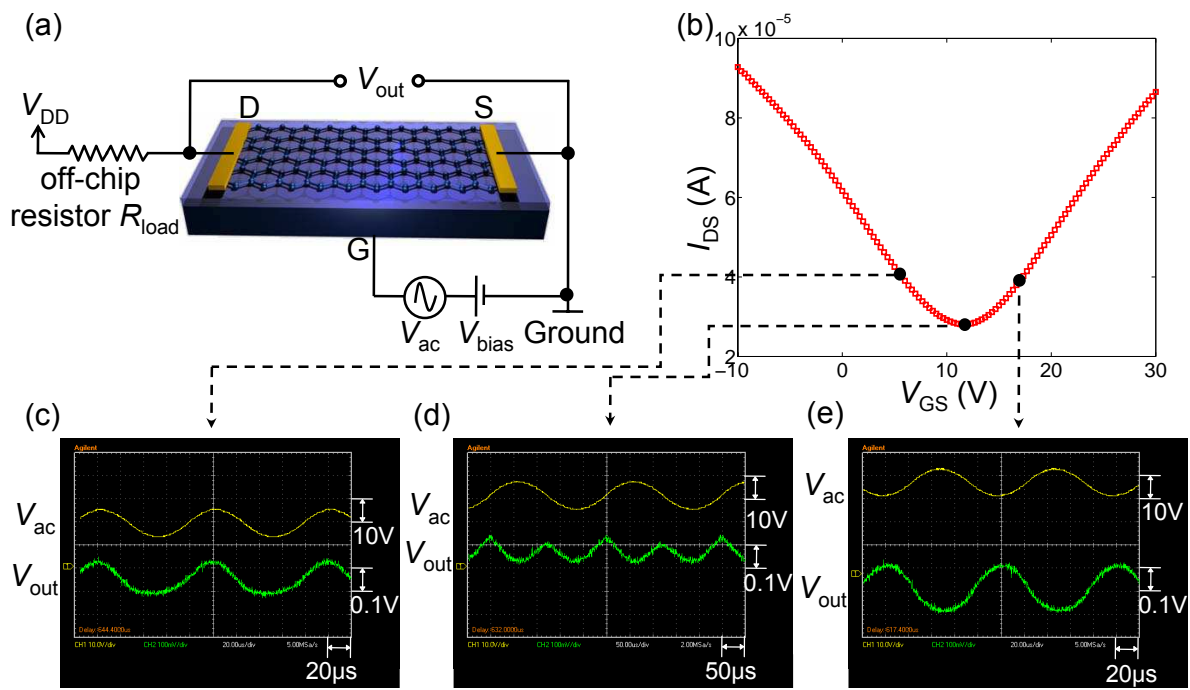}
\caption{
(a)~The schematic for the triple-mode single-transistor graphene amplifier
based on an off-chip resistor $R_\text{load}$.
(b)~The $I_\T{DS}$-$V_\T{GS}$ characteristics of the graphene transistor. The three
dots represent three representative bias voltages for the three different modes of operation.
From the left to the right, for the three bias voltages, the amplifier is configured
in the common-drain mode, the frequency multiplication mode, and the common-source mode, respectively.
(c)~The AC coupled input and output signals when the amplifier is biased at the left branch
of the ambipolar curve. In this configuration, the amplifier is in the common-drain mode,
and the output signal has the same frequency and phase as the input signal.
(d)~The AC coupled input and output signals when the amplifier is biased at the Dirac point.
In this configuration, the amplifier is in the frequency multiplication mode,
and the frequency of the output signal is doubled as compared to that of the input signal.
(e)~The AC coupled input and output signals when the amplifier is biased at the right branch
of the ambipolar curve. In this configuration, the amplifier is in the common-source mode,
and the output signal has the same frequency but a 180$^\circ$
phase shift as compared to the input signal.
}
\label{fig:OPC}
\end{figure}

\hl{Triple-mode amplifier circuit setup.} 
The schematic of the graphene amplifier is shown in~\ref{fig:OPC}(a).
$V_\T{bias}$ is a fixed DC voltage and $V_\T{ac}$ is a small sinusoidal AC signal.
The input and output voltages of the 
triple-mode single-transistor graphene amplifier
are measured using an oscilloscope (Agilent DSO3102A).
$V_\T{bias}$ and $V_\T{DD}$ are applied using a power supply (Kepco ABC 40-0.5),
and $V_\T{ac}$ is applied using a signal generator (GM Instek GFG 8020H).

\begin{acknowledgement}
The work at Rice University was supported by NSF grant CCF-0916636.
The work at the University of California-Riverside was supported
by the DARPA-SRC Focus
Center Research Program (FCRP) through its Center on Functional Engineered Nano
Architectonics (FENA).
\end{acknowledgement}

\pagebreak

\pagebreak
% \footnotetext[4]{Figure and text to summarize the small-signal hybrid-$\pi$ model and
% small-signal circuit analysis.}

\pagebreak

\pagebreak

\pagebreak

% \pagebreak
% \begin{figure}
% \centering
% \includegraphics[width=.8\columnwidth]{enhanced}
% \caption{Circuit structure utilizing feedback and differential
% output to reduce the performance difference between
% different modes of operation.}
% \label{fig:enhanced}
% \end{figure}

%
\bibliography{ambipolar}

\end{document}